\pdfoutput=1

\documentclass[11pt]{article}

\usepackage{ACL2023}

\usepackage{times}
\usepackage{latexsym}

\usepackage{amsmath}
\usepackage{graphicx}
\usepackage{float}
\usepackage[caption = false]{subfig}
\usepackage{amsmath,amssymb}

\usepackage{amsmath}
\usepackage{float}
\usepackage[caption = false]{subfig}
\usepackage{algorithm}
\usepackage{algorithmic}
\usepackage{amsmath,amssymb}
\usepackage{dsfont}
\usepackage{multirow}
\usepackage{enumitem}
\usepackage{makecell}
\usepackage{tabularray}

\usepackage[T1]{fontenc}

\usepackage[utf8]{inputenc}

\usepackage{microtype}

\usepackage{inconsolata}

\title{Agent-S: LLM Agentic workflow to automate Standard Operating Procedures}

\author{Mandar Kulkarni\\
  Flipkart Data Science \\ Seattle, Washington, USA \\
   }

\begin{document}
\maketitle
\begin{abstract}
AI agents using Large Language Models (LLMs) as foundations have shown promise in solving complex real-world tasks.
In this paper, we propose an LLM-based agentic workflow for automating Standard Operating Procedures (SOP). For customer care operations, an SOP defines a logical step-by-step process for human agents to resolve customer issues. We observe that any step in the SOP can be categorized as user interaction or API call, while the logical flow in the SOP defines the navigation. We use LLMs augmented with memory and environments (API tools, user interface, external knowledge source) for SOP automation. Our agentic architecture consists of three task-specific LLMs, a Global Action Repository (GAR), execution memory, and multiple environments. SOP workflow is written as a simple logical block of text. Based on the current execution memory and the SOP, the agent chooses the action to execute; it interacts with an appropriate environment (user/API) to collect observations and feedback, which are, in turn, inputted to memory to decide the next action. The agent is designed to be fault-tolerant, where it dynamically decides to repeat an action or seek input from an external knowledge source. We demonstrate the efficacy of the proposed agent on the three SOPs from the e-commerce seller domain. The experimental results validate the agent's performance under complex real-world scenarios. 

\end{abstract}

\section{Introduction}
Large Language Models (LLMs) have demonstrated great potential in logical reasoning. However, LLMs fall short when tackling more sophisticated tasks that involve interaction with multiple environments. AI agents built on LLMs can control the path to solving a more complex task.



\begin{figure*}[htb]
  \centering
  \includegraphics[scale= 0.25]{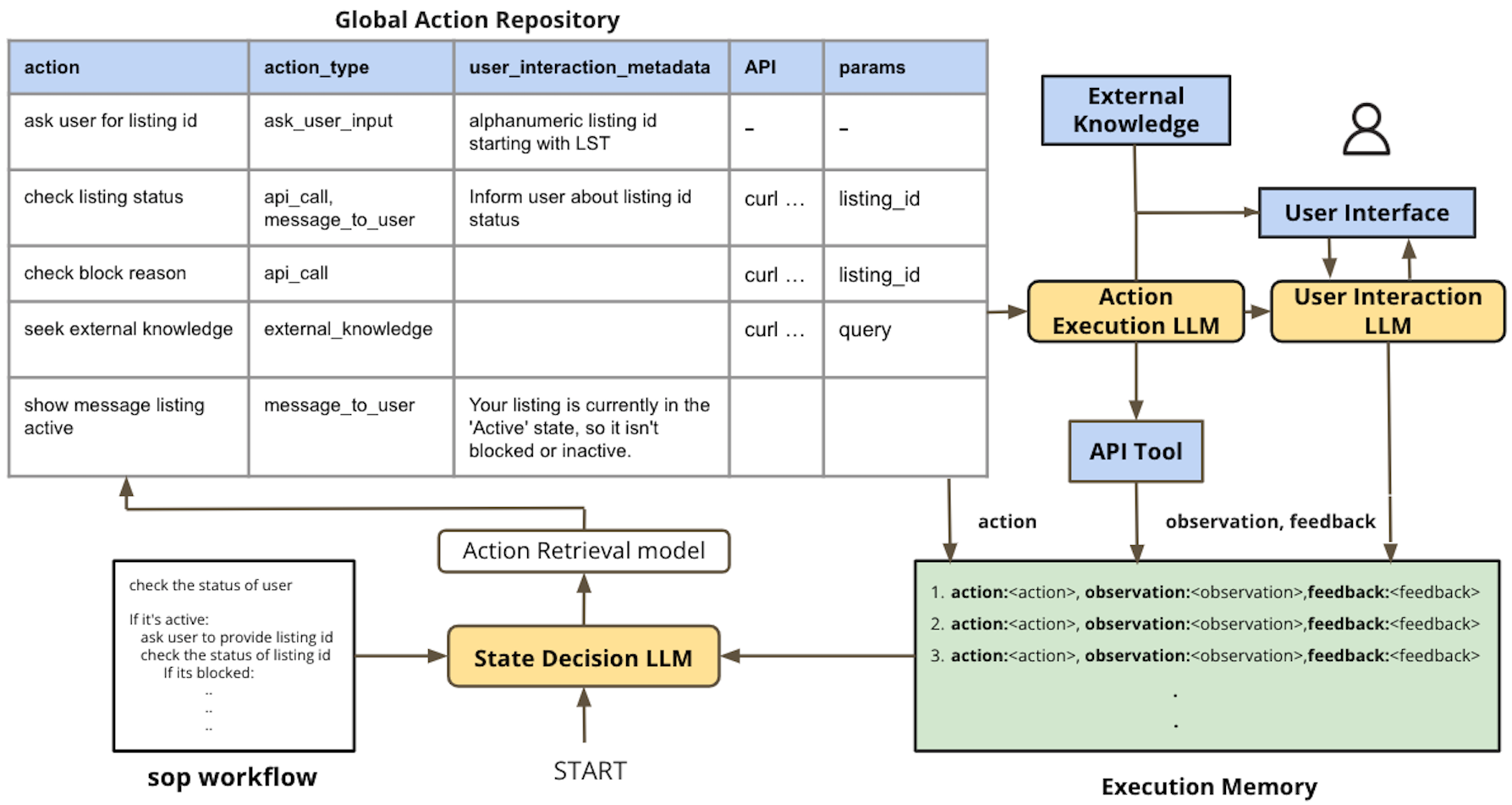}
  \caption{Proposed architecture of the SOP Agentic workflow.}
  \label{fig:sop_agent}
\end{figure*}\label{fig:arch}

A Standard Operating Procedure (SOP) defines a systematic workflow for human agents. It is a pre-defined step-by-step process designed to help human agents resolve recurring customer issues. When a consumer reaches out to customer care, a human agent identifies the issue (i.e., intent) and follows the sequence of steps from the SOP. 
While analyzing the SOPs and customer-human agent interactions from the historical e-commerce data, we observed that any step in the SOP can be categorized as follows.
\textbf{1. User interaction}: In this, human agents interact with the consumers. They ask the user relevant questions, interpret the answers, collect relevant information (e.g., product/listing ID, etc), and convey messages. \textbf{2. Status check}: Based on the information collected from the user, human agents check the status of product/listing ID on the dashboard (where data is populated through backend API calls).
Based on the output of the user interaction or status check, an agent decides the next action as per the SOP. 
If the current step in the SOP is executed successfully, the agent proceeds to the next step. The next action would again involve a user interaction or status check. If the current step has failed, an agent attempts to complete it either by repeating the current step or any of the previous dependent steps e.g., if during a status check, the agent realizes that the user has provided an invalid listing ID, the agent would repeat the step of asking user for the listing ID. 

In this paper, we propose an LLM based agentic workflow for executing SOPs to automate a human agent process for customer care support.
Both user interaction and status check steps can be automated using LLMs augmented with external environments such as API tools and user interface. The LLMs can generate the question to the user, it can interpret the user inputs, extract relevant information, it can generate status messages and acknowledgments. The status check step can be automated with the combination of API tools and LLMs. An API tool can help retrieve the information from the database, and an LLM can effectively utilize it to make decisions. Equipped with these capabilities, LLMs-based agents can, therefore, help to navigate the entire SOP flow in an automated manner. Technically, the SOP can be thought of as a Directed Acyclic Graph (DAG), where each node represents the sop step (state), and the edges represent the possible branches from the current state.

Fig. \ref{fig:arch} shows the proposed architecture of the SOP agent. The architecture consists of three LLM components, a State-Decision-LLM: to decide the next action (state) in the SOP, an Action-Execution-LLM: to help execute the currently selected action and a User-Interaction-LLM: to interpret user inputs and provide acknowledgments.
The inputs to the state decision LLM are the sop workflow and execution memory. In our approach, the SOP workflow can simply be written as a logical block of text with indents depicting sub-flows. Table \ref{tab:listing} shows an SOP for listing blocked issue from the e-commerce seller domain. Execution memory records the history of execution where each entry contains the selected action and its corresponding observation and feedback. The observation and feedback are obtained from the environment with which the agent interacts to execute the action. In the case of an API call, observation contains the (parsed) API response or error message, while if asking for the user's input, the observation contains the spell corrected user's input from the user-interaction-LLM. Feedback indicates whether the current action has been executed successfully or not. The Global Action Repository (GAR) table contains a set of possible actions and the corresponding required information to execute the actions.
Using sop workflow and current execution memory, the state-decision-LLM decides the next action to be executed. GAR provides the required info to action-execution-LLM to execute the action. Depending on the type of the action, the action-execution-LLM interacts with API tool or user interface. The observation and feedback are generated and inputted to the execution memory, and the state-decision- LLM predicts the next action. The process is repeated till the termination.

From historical user-human interactions data, we also noticed that users often ask questions/doubts regarding the requested information, e.g., when asked to provide a listing ID, they typically ask where/how to find their listing ID. We incorporate an external knowledge source in the architecture to seamlessly answer the user queries within the SOP flow. We use an in-house Retrieval Augmented Generation (RAG) system with domain documents to get the answer. When the agent identifies that the user is asking a question, it formulates a search query, fetches an answer from the RAG, and continues the SOP flow from where it branched.

To validate the effectiveness of the agent, we extensively experimented with three SOPs from the seller domain: why the listing is blocked, want to update the email ID, and why brand approval is rejected. 
We evaluated the SOP agent on synthetic and live chat sessions.
To mimic real world scenario, we synthetically generated a large variation of user inputs, all possible API responses and failure scenarios and verified the robustness of the agent. 
After a successful synthetic evaluation, we conducted in-house experiments with live chat sessions. The experimental results demonstrated that the LLM agent can effectively navigate the SOP flow. The agent is scheduled to go live for all sellers after engineering integration. We provide all LLM prompts in the Appendix for replicability.

\section{LLM Agentic workflow for SOP automation}
Fig. \ref{fig:arch} shows the proposed architecture. The architecture has the following main components.

\subsection{Global Action Repository (GAR)}

We maintain a Global Action Repository (GAR) as a common repository of all possible actions and the corresponding metadata. We refer to it as "Global" because this is a shared repository across all SOPs. GAR, in our case, is a table consisting of the following columns.
\begin{itemize}[noitemsep,nolistsep]
\item \textbf{action}: The action identifier for the SOP step that needs to be executed 
\item \textbf{action\_type}: It defines the type of the action. An action type can include one or many of the following: api\_call, ask\_user\_input, message\_to\_user, or external\_knowledge. 
\item \textbf{user\_interaction\_metadata}: For the user interaction action types (ask\_user\_input or message\_to\_user), it contains the inputs expected from users or instruction/message to be conveyed.
\item \textbf{API}: If the action type is api\_call, it contains the corresponding API endpoints e.g., listing status check API
\item \textbf{params}: It enlists the input parameters required for the API e.g., listing\_id would be needed for listing status check API
\end{itemize}

\subsection{State Decision LLM}
The most important LLM module in our architecture is the State Decision LLM. We will refer to this LLM as a state LLM. The task of state LLM is to decide the next action the agent should execute, i.e., decide the state of the agent. We use a chain of thought prompting for the state LLM. The state LLM has two inputs:\\
1. Pre-defined SOP workflow\\
2. Execution memory\\ 
\vspace{-0.5cm}
\subsubsection{SOP workflow}

The SOP workflow defines the control flow over the steps and the outcomes. An advantage of the proposed agent is that the SOP flow can be written as a logical block of text with indents specifying the sub-flow. We do not enforce any specific structure/schema for the SOP. Table \ref{tab:listing} shows the workflow for listing blocked SOP defined in terms of actions and outcomes (observations).

\begin{table}[h]
    \centering
    \captionsetup{justification=centering, margin=5mm}
    \begin{tabular}{p{7cm}}
        
        
    \hline
        check user status
        \vspace{0.02cm}
        
        if its onboarding:
        
        \hspace{0.3cm} show message onboarding
        
        \hspace{0.3cm}   terminate the flow

\vspace{0.02cm}

if its active or on-hold:
    
    \hspace{0.3cm} ask user to provide listing id
    
    \hspace{0.3cm} check listing id status

    \vspace{0.02cm}
    
    \hspace{0.3cm} if its inactive:
        
        \hspace{0.8cm} show message listing inactive
        
        \hspace{0.8cm} terminate the flow

    \vspace{0.05cm}
    
    \hspace{0.3cm} if its active:
        
        \hspace{0.8cm} show message active listing
        
        \hspace{0.8cm} terminate the flow

    \vspace{0.02cm}
    
    \hspace{0.3cm} if its blocked:
                
        \hspace{0.8cm} check block reason
        \vspace{0.02cm}
        
        \hspace{0.8cm} if block reason is seller state change:
            
            \hspace{1.2cm} show message seller state change
            
            \hspace{1.2cm} terminate the flow
        
        \vspace{0.02cm}
        
        \hspace{0.8cm} else:
        
            \hspace{1.2cm} check if listing can be reactivated
            
            \hspace{1.4cm} if yes:
                
                \hspace{1.6cm} show message reactivation 
                
                \hspace{1.6cm} create ticket
                
                \hspace{1.6cm} terminate the flow
            
            \hspace{1.2cm} if no:
                
                \hspace{1.6cm} check reason code and inform user
                
                \hspace{1.6cm} terminate the flow \\ \hline
            
    \vspace{0.1cm}

    \end{tabular}
    \vspace{2mm}
    \caption{Listing blocked SOP workflow}
    \label{tab:listing}
\end{table}

\subsubsection{Execution Memory}

We design a textual execution memory that maintains the history of the execution. 
Each entry in the execution memory consists of three components: action,  observation, and feedback. \textbf{1. action}: the current selected action from the GAR, 
\textbf{2. observation}: An observation is a response an agent receives when it interacts with the environment. Based on the type of selected action, the agent interacts either with the user interface or API tool and receives the observation.  
\textbf{3. feedback}: It indicates whether the current action was executed successfully or not. In our case, feedback is assigned a binary value: success or fail.
 
As shown in Fig. \ref{fig:arch}, the (observation feedback) pair is inputted to the execution memory either from the API tool or from the User- -interface-LLM. If the selected action type is ask\_user\_input, the observation is set to the spell-corrected user input obtained from user- interface-LLM. The feedback is set to success or fail based on whether the input validation is successful or not. The user-interaction-LLM performs input validation as explained in section \ref{sec:ui}.
If the selected action type is api\_call, the observation is set to the (parsed) API response. The API response parsing is implemented within the API tool. Feedback indicates the success of the API call. If the API returns an error, e.g., API call failed or invalid ID, the observation is set to the error message, and the feedback is set to fail. If the API call fetches the status successfully, feedback is set to success. For external\_knowledge and message\_to\_user action types, observation is set to a standard text as "done" and feedback is set to success/fail based on whether the action was executed successfully.

\subsection{Action Retrieval model}
Since the state LLM is a generative model, we use an embedding model to identify the action from the set of possible actions from GAR. We get the vector representation of all actions from GAR using the retrieval model. When the state LLM outputs the next action, it is encoded using the same model. We then perform a cosine similarity-based search for the actions in GAR, and the best match is considered as the selected action. Due to semantic search,
the names of the action identifiers from GAR need not exactly match the ones generated by the state LLM. This allows more variations in writing SOPs. We use a pre-trained e5-base-v2 \cite{wang2022text} as the action retrieval model.

\subsection{Action Execution LLM}\label{sec:ae}
The task of Action Execution LLM is to generate the data required for executing the selected action. We refer to this LLM as an action LLM. 
It is a multi-purpose LLM that is prompted to perform the task according to the type of action.
The GAR provides the following inputs to the action LLM: (selected) action, its action type, and task context. 
Based on the action type, the action LLM performs following tasks:
\textbf{1.} generate a question to the user: If the selected action type is ask\_user\_input, the LLM is prompted to generate the question to the user based on the action. The generated question is passed to the user interface. Since the action name itself is used, the task context is empty for this task.
\textbf{2.} extract parameters for API call: If the selected action type is api\_call, the LLM is prompted to extract the parameters to be passed to the API from the task context. The task context, in this case, contains the required list of input parameters and the slots (entities) identified till now from the User Interface LLM. As explained in section \ref{sec:ui}, the user interface LLM assigns generic names to the slots, and the task of action LLM is to map the slots to the required input parameters. The extracted input parameters, along with the API endpoint, are passed to the API tool.
\textbf{3.} generate a message to the user: 
If the selected action type is message\_to\_user, the LLM is prompted to generate the message to the user. In this case, the task context contains the message to be conveyed obtained from the user\_interaction\_metadata. The generated message is passed to the user interface.
\textbf{4.} generate the search query: 
If the selected action type is external\_knowledge, the LLM is prompted to generate the search like query to fetch the answer from the external knowledge source. Task context, in this case, contains the execution memory so that the LLM has the context about the action and query asked by the user (i.e., observation).


\vspace{-0.2cm}
\subsection{User Interaction LLM}\label{sec:ui}
The task of user interaction LLM is to interpret user inputs, extract entities, and provide acknowledgment messages. 
We refer to this LLM as the user LLM. We provide the following inputs to the LLM: the question generated from the action LLM, the user's reply to the question, and the input expected from the user (obtained from GAR user\_interaction\_metadata). 
The LLM is prompted to perform the following tasks.
\textbf{1.} input validation: The task is to perform a sanity check and input validation. The LLM is prompted to validate the user input against the expected input (from GAR). If the user input is as expected, the input validation is set to success else, it is set to fail. 
\textbf{2.} extract slots: The task is to extract the entities (slots) from the user inputs to be used for subsequent API calls, e.g., listing ID value provided by the user. The LLM is prompted to output slots as key-value pairs. The user LLM assigns generic names to the slot keys, and the action LLM maps the generic slot names to the required params as explained in section \ref{sec:ae}.
\textbf{3.} generate acknowledgment: The task is to generate a brief acknowledgment message to the user based on the input validation status
\textbf{4.} spell correction: The task is to do a spell correction on the user input (to be inputted as observation to execution memory)
\vspace{-0.2cm}

\section{Details of the agent execution}

At the start, when the execution memory is empty, the state LLM is prompted to select the first step from the SOP. The generated action is vector encoded using an action retrieval model, and the best matching action is selected from GAR using cosine similarity. The selected action and its corresponding row entries are passed to the action LLM. Based on the action type, action LLM generates a question/message to the user, extracts the parameters for the API call, or formulates a search query for the external knowledge base. If the action type is ask\_user\_input, the user receives the question (generated from action LLM). The user's reply is inputted to the user LLM. The user LLM performs input validation and slot extraction and generates appropriate acknowledgment messages. The user LLM provides the observation and feedback to the execution memory. If the selected action type is api\_call, the API tool generates the observation and feedback and adds it to the execution memory. The state LLM gets an updated execution memory input, and it then decides the next logical step to be executed. 
If the current action failed due to some error, the failure reason is captured in the observation (e.g., API call failure, invalid ID), and the LLM is prompted to \emph{dynamically} decide the logical previous action to be repeated in order to proceed. Note that the state LLM chooses the most apt previous action based on logical reasoning. Interestingly, during experiments for listing blocked SOP, we note that when the API call fails, the LLM decides to repeat the failed API call action; however, when the API call returns an invalid ID as the error, the LLM decides to repeat the asking listing ID to user step and continues execution from that point. 
Table \ref{table:laf} and \ref{table:lai} in the Appendix show the behavior of the agent for failure cases depicting the content of execution memory, outputs of state and action LLM, and generated observation and feedback. 

In the happy scenario with no failures, the flow is executed until the termination step mentioned in the SOP. However, it's possible for the agent to get stuck in the loop/cycle e.g., even if the user has provided an invalid listing ID multiple times, the most logical action for the state LLM is still to ask for the listing ID. To avoid such scenarios, we maintain the count of the number of times the action is repeated.
If the particular action is repeated more than two times, we terminate the agent flow with a grace message to the user.

Many times, users have doubts and ask questions when prompted to provide some info e.g., when asked to provide the listing ID, we see that sellers have questions as to how/where to find it. 
The expected agent behavior, in this case, is to show users the required answer and resume the SOP execution.
We prompt state LLM to output the state as external\_knowledge when the user asks a question. The action LLM constructs a search query using execution memory as context. We display the answer obtained from the external knowledge base (i.e., from RAG on the set of domain documents). The state LLM then continues the execution by asking the user to provide the required info again. Table \ref{table:uhq} in Appendix shows the behavior of the agent when user asks a doubt within SOP flow. 

\section{Evaluation}

We performed extensive evaluation of the proposed agentic approach with synthetic as well live chat sessions on three high volume SOPs from the e-commerce seller domain: why my listing is blocked, why my brand approval is rejected (Appendix section \ref{tab:bas}) and want to update email id (Appendix section \ref{tab:eus}). 

We created synthetic chat sessions to test the agent's robustness under all possible scenarios. The synthetic user inputs are generated to mimic the variations observed in the real-world data. These contain valid inputs,  chitchat and gibberish inputs, users asking doubts/queries, a large set of invalid inputs, and inputs with invalid formats (e.g., email ID without domain name). The synthetic data for API responses is generated to cover all possible API outputs and failure cases, e.g., API call failures, API returning errors as invalid inputs, etc. A chat session is created by randomly sampling a user input and API response for corresponding steps, thus creating a different traversal of the SOP. For three SOPs, we created 220 synthetic sessions containing 1221 different states. 
The agent is provided with the synthetic input at each turn, and the state predicted by the state LLM is recorded. 
We manually evaluated state LLM accuracy as follows: If the state LLM predicts the expected step as the next action, it is labeled correct; otherwise, it is labeled incorrect. If the state LLM continues the SOP flow even if one of the previous steps has failed,  the entire session from the point of failure is marked as incorrect, e.g., if the state LLM continues to the next action even though the user has provided an invalid listing ID, the entire session is after the erroneous step is marked as incorrect. 

We evaluated two public LLMs for the state LLM, ChatGPT-3.5 (gpt-35-turbo-0613) and ChatGPT-4o-mini (gpt-4o-mini). Table \ref{tab:st} compares the accuracy of the two LLMs. We see that ChatGPT-4o-mini works significantly better than ChatGPT-3.5. In the errors, we observe that ChatGPT-3.5 often chooses incorrect action for valid and invalid inputs, while ChatGPT-4o-mini predicts the next action accurately. The result indicates that an LLM with good logical reasoning capability is necessary for good performance.

\begin{table}[H]
\centering
\begin{tabular}{|l|l|}
\hline
\textbf{LLM} & \textbf{Accuracy} \\
\hline
GPT-3.5 & 0.565\\
\hline
GPT-4o-mini &  \textbf{0.978} \\
\hline

\end{tabular}
\caption{Evaluation of state LLM accuracy}
\label{tab:st}

\end{table}

Next, the action LLM accuracy is evaluated only for the cases where the state LLM decision is correct. This is because if the state LLM predicts an incorrect action, the action LLM would not receive an appropriate task context and thus would fail. We manually evaluated the action LLM accuracy of the three tasks: generating questions for users, extracting parameters for API calls, and generating search queries for the external knowledge base. Table \ref{tab:ac} shows the comparison result. 
Both models provide high accuracy for question generation and parameter extraction tasks, which is expected given their simplistic nature. Note that the accuracy of action LLM for parameter extraction tasks is directly dependent on the accuracy of user LLM for slot extraction.
If user LLM fails to capture entities (slots) from the user's reply, 
action LLM will fail to predict params. Hence, the accuracy of action LLM indirectly reflects the accuracy of the user LLM.
For search query generation, ChatGPT-4o-mini performs significantly better than ChatGPT-3.5.  

\begin{table}[H]
\centering
\begin{tabular}{|l|l|l|}
\hline
\textbf{Task} & \textbf{gpt-3.5} & \textbf{gpt4o-mini}\\
\hline
question generation  & 1. & 1. \\
\hline
parameter extraction & 0.981 &  1.\\
\hline
search query generation & 0.44 &  0.951\\
\hline

\end{tabular}
\caption{Evaluation of action LLM accuracy}
\label{tab:ac}

\end{table}

We also tested the agent's performance with in-house live chat sessions. We observed good accuracy for the session success. The agent is scheduled to go live for all sellers post Engg. integration. 

\section{Related works}
LLM-based agents are being experimented in different applications. Some of them include web navigation \cite{abuelsaad2024agenteautonomouswebnavigation}, travel planning \cite{xie2024travelplannerbenchmarkrealworldplanning}, video understanding \cite{zhang2024omagentmultimodalagentframework}, biomedical discovery  \cite{gao2024empoweringbiomedicaldiscoveryai}, code documentation  \cite{luo2024repoagentllmpoweredopensourceframework}, knowledge graph reasoning \cite{jiang2024kgagentefficientautonomousagent}, knowledge base question answering \cite{zong2024triadframeworkleveragingmultirole}.
Memory is a crucial component of agents. Zhang et al. \cite{zhang2024surveymemorymechanismlarge} provided an extensive survey of memory methods for agents.


Wu et al. \cite{wu2024stateflowenhancingllmtasksolving} proposed an LLM-based state machine paradigm for complex task-solving processes. The state transitions are controlled by the LLM or by heuristic rules. Each state can perform a series of actions augmented with external tools. 
In the state machine approach, an LLM prompt needs to be developed for each state to guide the state transitions. Our approach has an advantage in that the single-state LLM prompt can predict the actions across all SOPs. Therefore, our approach offers more flexibility in terms of SOPs and error handling.

\section{Conclusion}
In this paper, we proposed an agentic workflow for SOP automation. The proposed architecture consisted of task-specific LLMs, execution memory, and a shared action repository across all SOPs. The SOPs can be written as simple logical blocks of text without stringent formatting requirements. Experimental results with synthetic and live chat sessions demonstrated the efficacy of the proposed agent. The results indicated that an LLM with good reasoning capability is necessary. Even though we have experimented with SOPs, the proposed framework can be used to automate workflows represented as Directed Acyclic Graphs (DAG).

\bibliography{custom}
\bibliographystyle{acl_natbib}

\appendix
\clearpage

\section{Appendix}

\subsection{State LLM prompt}

\begin{table}[h]
    \centering
    \captionsetup{justification=centering, margin=5mm}
    \begin{tabular}{p{15cm}}
        
        
        \hline
I want you to act as the action decision agent for the workflow automation task. 

\vspace{0.1cm}

You will be provided with the following information.

1. Workflow\\
2. Execution Memory

\vspace{0.1cm}
Workflow consists of a logical sequence of actions. 
Execution Memory consists of the history of actions, observations and feedback.

\vspace{0.1cm}

Your task is to decide the next action based on the workflow and execution memory. 

\vspace{0.1cm}

*** If the execution memory is empty, output the first action from the workflow.

\vspace{0.1cm}

*** If the feedback for the current entry in execution memory mentions success, output the next action as per the logic shown in the workflow.

\vspace{0.1cm}

*** If the feedback for the current entry in execution memory mentions fail, decide the next action as follows.

    -- If observation indicates that the user wants to go back to any of the previous actions, perform a semantic search in the current execution memory and find the relevant action to help the user. Output the action as the next action.
    Post this you must continue the workflow from where it was broken.

    -- If the observation clearly indicates that the user has a question or a query, output the action as seek external knowledge.
    If feedback for the seek external knowledge action step is success, output the previous valid action from the Execution Memory as the next action.    

\vspace{0.1cm}

*** If the feedback for the last entry in execution memory mentions fail and observation does not clearly indicate any of the above scenarios, decide the next action as follows.
Carefully evaluate the inter-dependence of the current failed action on the previous actions in the execution memory and select the most logical previous action that needs to be repeated. Output it as the next action. 

\vspace{0.2cm}

\#\#\# Workflow:
<sop\_workflow>
\#\#\#

\vspace{0.2cm}

\#\#\# Execution Memory:
<execution\_memory>    
\#\#\#

\vspace{0.2cm}

Think step by step and output your thinking as thought.   

Generate all the responses in the JSON format without any deviation. Output JSON should have keys "thought, "next\_action". \\ \hline

        \end{tabular}
    \vspace{2mm}
    \label{tab:sllm}
\end{table}

\clearpage
\subsection{Action LLM prompt}

\begin{table}[h]
    \centering
    \captionsetup{justification=centering, margin=5mm}
    \begin{tabular}{p{15cm}}

    \hline
I want you to act as the action execution agent for the workflow automation task. 
\vspace{0.1cm}
You will be provided with the following information.

1. Action in the workflow\\
2. Action type\\
3. Action context

\vspace{0.1cm}

Your task is to generate data to execute an action as per the action, action type and action context.

\vspace{0.1cm}

1. If action type includes ask\_user\_input, your task is to generate a polite question to the user using the action. Output the question as user\_interaction.

\vspace{0.1cm}

2. If action type includes api\_call, your task is to extract and assign a correct value to each of the required param using the action context. Output the required params and its values.

\vspace{0.1cm}

3. If action type includes external\_knowledge, your task is formulate a short search like query from the user's question/query provided in the action context. Output the search query as search\_query.

\vspace{0.1cm}

4. If action type includes message\_to\_user, your task is to generate the response to the user as shown in the action context. For failure case, inform user that you are retrying the <action>. Output the response as user\_interaction.

\vspace{0.1cm}

\#\#\# Action: 
<action>

\#\#\# Action type: 
<action\_type>

\#\#\# Action context: 
<action\_context>

\vspace{0.1cm}

Think step by step and output your thinking as thought.

Generate all the responses in the JSON format without any deviation. Output JSON should have keys "thought, "user\_interaction", "params", "search\_query". \\ \hline

    \end{tabular}
    \vspace{2mm}
    \label{tab:sllm}
\end{table}

\clearpage

\subsection{User LLM prompt}
\begin{table}[h]
    \centering
    \captionsetup{justification=centering, margin=5mm}
    \begin{tabular}{p{15cm}}

    \hline
I want you to act as the user interaction agent for the workflow automation task. 

\vspace{0.1cm}

You will be provided with the following information.

\vspace{0.1cm}

1. Question asked to the user\\
2. User's reply\\
3. Condition

\vspace{0.1cm}

Your tasks are as follows.

\vspace{0.1cm}

1. Verify if the user's reply satisfies the condition. 
   
If yes, set input\_validation field as success. Otherwise set it as fail. 

\vspace{0.1cm}

2. Extract all the entities from user's reply and output the slots with key and value per entity. Assign a distinctive name to the key as per the question for easy identification.

\vspace{0.1cm}

3. Generate a response to the user as follows. 
    
    If input\_validation is success, provide a one-line acknowledgment message.
        
    If input\_validation field is fail:  

\vspace{0.1cm}
        
    ** If User's reply clearly shows a question or a query, output the message that you are working on it and politely ask user to wait. 
    
    ** If User's reply is not a question or a query, provide a one-line acknowledgment message.

\vspace{0.1cm}

\#\#\# Question asked to the user: 
<question>
    
\#\#\# User's reply:
<user\_reply>

\#\#\# Condition: User's reply which indicates or includes <expected\_format>

\vspace{0.1cm}

Think step by step and output your thinking as thought.

Generate all the responses in the JSON format without any deviation. Output JSON should have keys "thought, "input\_validation", "user\_response", "slots". \\ \hline

    \end{tabular}
    \vspace{2mm}
    \label{tab:sllm}
\end{table}

\clearpage

\begin{table*}[t!]
  \caption{Agent behavior when checking listing ID status API call fails. The API tool returns the observation that the API call has failed, and feedback is set to fail. The agent has logically decided to repeat the failed check listing ID status API call again.}
  \label{table:laf}
  \centering
  \begin{tblr}{
    colspec = {|X[4,m,c]|X[3,m,c]|X[3,m,c]|X[5,m,c]|X[3,m,c]|},
    cell{2,5,8,11,14,17,20,23,26}{2} = {r=1}{c}, 
    cell{2,5,8,11,14,17,20,23,26}{2} = {r=1}{c}, 
  }
  \hline
    Execution memory &  state LLM output & action LLM output & User's reply/message to user & observation,\ feedback \\
  
    \hline
    1. action:check user status, \ observation:active, \ feedback:success
    2. action:ask user to provide listing id, \ observation: \ LSTFYDF12G,\ feedback:success  & check listing id status & listing\_id: 'LSTFYDF12G' & The status of your listing ID (LSTFYDF12G) could not be retrieved due to an error. I am retrying the check for you. & observation: api call failed \ feedback: fail \\

    \hline

    1. action:check user status, \ observation:active, \ feedback:success
    2. action:ask user to provide listing id, \ observation: \ LSTFYDF12G,\ feedback:success 3. action:check listing id status, \ observation: \ api call failed,\ feedback:fail & check listing id status & listing\_id: 'LSTFYDF12G' & The status of the listing ID 'LSTFYDF12G' is Active.  & observation: done \ feedback: success \\

    \hline

  \end{tblr}
\end{table*}

\begin{table*}[h]
  \caption{Agent behavior when checking listing ID status API returns an invalid ID as an error. The API tool returns the observation as the invalid ID, and feedback is set to fail. The agent has logically decided to repeat the action of asking the user to provide the listing ID step again.}
  \label{table:lai}
  \centering
  \begin{tblr}{
    colspec = {|X[4,m,c]|X[3,m,c]|X[3,m,c]|X[5,m,c]|X[3,m,c]|},
    cell{2,5,8,11,14,17,20,23,26}{2} = {r=1}{c}, 
    cell{2,5,8,11,14,17,20,23,26}{2} = {r=1}{c}, 
  }
  \hline
    Execution memory &  state LLM output & action LLM output & User's reply/message to user & observation,\ feedback \\
  
    \hline
    1. action:check user status, \ observation:active, \ feedback:success
    2. action:ask user to provide listing id, \ observation: \ LST1234,\ feedback:success  & check listing id status & listing\_id: 'LST1234' & The status of the listing ID 'LST1234' is invalid. I am retrying the check listing id status. & observation: invalid listing id \ feedback: fail \\

    \hline

    1. action:check user status, \ observation:active, \ feedback:success
    2. action:ask user to provide listing id, \ observation: \ LST1234,\ feedback:success 3. action:check listing id status, \ observation: \ invalid listing id,\ feedback:fail & ask user to provide listing id & Could you please provide the listing ID? & LSTFYDF12G & observation: LSTFYDF12G \ feedback: success \\

    \hline

  \end{tblr}
\end{table*}


\begin{table*}[!t]
  \caption{An example of agent behavior is when the user asks a question within the SOP flow. When asked to provide the listing ID, the user asked the doubt as, 'how to find it'. In this case, the user LLM returns the observation as the user inputted text, and feedback is set to fail since an alphanumeric listing ID was expected. The state LLM has predicted seek external knowledge as the next action to be executed. The action LLM correctly forms a search query (How to find my listing ID), considering execution memory as the task context. Note that it replaced the contextual word 'it' in the query with 'listing ID.' We use this search query to get the answer from the in-house RAG system. After completing this, the agent again asks the user to provide the listing ID. This way, we can provide a seamless experience to users.}
  \label{table:uhq}
  \centering
  \begin{tblr}{
    colspec = {|X[4,m,c]|X[3,m,c]|X[3,m,c]|X[5,m,c]|X[3,m,c]|},
    cell{2,5,8,11,14,17,20,23,26}{2} = {r=1}{c}, 
    cell{2,5,8,11,14,17,20,23,26}{2} = {r=1}{c}, 
  }
  \hline
    Execution memory &  state LLM output & action LLM output & User's reply/message to user & observation,\ feedback \\
  
    \hline
    1. action: check user status,\ observation:active,\ feedback: success  & ask user to provide listing id & Could you please provide the listing ID? & how to find it  & observation: how to find it 
    \ feedback: fail \\
     
    \hline
    1. action: check user status,\ observation:active,\ \vspace{0.1cm} feedback: success  \  2. action: ask user to provide listing id,\ observation:how to find it,\ feedback: fail  & seek external knowledge & search query: How to find my listing ID? & To find your Listing ID, follow these steps: 1. Log into your Seller Portal 2. Under the 'Listings' tab, select 'My Listings' 3. Search for the product using FSN/Title/SKU ID 4. Click on the 'Edit Listing' against the FSN 5. On the right-hand side, click on 'Listing Information' 6. Under the 'Status Details', check the 'Listing Status'  & observation: done \ feedback: success  \\
     
    \hline
    1. action: check user status,\ observation:active, \ feedback: \vspace{0.1cm} success  \ 2.action: ask user to provide listing id,\ observation:how to find it,\ \vspace{0.1cm} feedback: fail \ 3. action: seek external knowledge ,\ observation:done,\ feedback: success & ask user to provide listing id & Could you please provide the listing ID? & my listing id is LSTHFKKFL  & observation: my lisint id is LSTHFKKFL \ feedback: success \\
     
    \hline
  
  \end{tblr}
\end{table*}

\clearpage
\subsection{Email update SOP}\label{tab:eus}

\begin{table}[h]
    \centering
    \captionsetup{justification=centering, margin=5mm}
    \begin{tabular}{p{15cm}}

    \hline
    check user status

        \vspace{0.1cm}

        if its on-hold or onboarding:
        
        \hspace{0.3cm} show message email update not possible
            
        \hspace{0.3cm} terminate the flow

\vspace{0.1cm}

if its active:

    \hspace{0.3cm} ask user about access to the old email  

    \hspace{0.3cm} if user has access:
    
    \hspace{0.8cm} ask user to provide old email
    
    \hspace{0.8cm} send otp and ask for otp received on old email
    
    \hspace{0.8cm} validate otp old email and inform user on validation status
    
    \hspace{0.8cm} ask user to provide new email
    
    \hspace{0.8cm} send otp and ask otp received on new email
    
    \hspace{0.8cm} validate otp new email and inform user on validation status
    
    \hspace{0.8cm} show message email updated
    
    \hspace{0.8cm} terminate the flow

\vspace{0.1cm}

\hspace{0.3cm} if user does not have access:
    
    \hspace{0.8cm} ask user to provide phone number
    
    \hspace{0.8cm} send otp and ask for otp received on phone number
    
    \hspace{0.8cm} validate otp phone number and inform user on validation status
    
    \hspace{0.8cm} ask user to provide new email
    
    \hspace{0.8cm} send otp and ask otp received on new email
    
    \hspace{0.8cm} validate otp new email and inform user on validation status
    
    \hspace{0.8cm} show message email updated
    
    \hspace{0.8cm} terminate the flow \\ \hline

    \end{tabular}
    \vspace{2mm}
    
\end{table}

\subsection{Brand approval SOP} \label{tab:bas}
\begin{table}[h]
    \centering
    \captionsetup{justification=centering, margin=5mm}
    \begin{tabular}{p{15cm}}
           
    \hline
   ask user to provide request id

check request id status
        \vspace{0.1cm}

        if approved:
        
        \hspace{0.3cm} show message brand approved
            
        \hspace{0.3cm} terminate the flow

        \vspace{0.1cm}

        if in-progress or disapproved:
        
        \hspace{0.3cm} if less than or equal to 72 hrs:
            
        \hspace{0.5cm} show message less than 72 hrs

        \hspace{0.5cm} terminate the flow

        \hspace{0.3cm} else:

        \hspace{0.5cm} create ticket brand approval

        \hspace{0.5cm} terminate the flow \\ \hline

    \end{tabular}
    \vspace{2mm}
    
\end{table}

\clearpage

\end{document}